\documentclass[]{emulateapj}

\begin{document}

\slugcomment{Accepted to ApJ}
\newcommand{\kms}{km s$^{-1}$}
\newcommand{\msun}{$M_{\odot}$}
\newcommand{\rsun}{$R_{\odot}$}

\title{ MMT Hypervelocity Star Survey. II.  Five New Unbound Stars }

\author{Warren R.\ Brown,
	Margaret J.\ Geller, and
	Scott J.\ Kenyon}

\affil{Smithsonian Astrophysical Observatory, 60 Garden St, Cambridge, MA 02138}

\shorttitle{ MMT Hypervelocity Star Survey II.}
\shortauthors{Brown, Geller, \& Kenyon}

\begin{abstract}

	We present the discovery of five new unbound hypervelocity stars (HVSs) in 
the outer Milky Way halo.  Using a conservative estimate of Galactic escape 
velocity, our targeted spectroscopic survey has now identified 16 unbound HVSs as 
well as a comparable number of HVSs ejected on bound trajectories. A Galactic center 
origin for the HVSs is supported by their unbound velocities, the observed number of 
unbound stars, their stellar nature, their ejection time distribution, and their 
Galactic latitude and longitude distribution.  Other proposed origins for the 
unbound HVSs, such as runaway ejections from the disk or dwarf galaxy tidal debris, 
cannot be reconciled with the observations. An intriguing result is the spatial 
anisotropy of HVSs on the sky, which possibly reflects an anisotropic potential in 
the central 10-100 pc region of the Galaxy.  Further progress requires measurement 
of the spatial distribution of HVSs over the southern sky.  Our survey also 
identifies seven B supergiants associated with known star-forming galaxies; the 
absence of B supergiants elsewhere in the survey implies there are no new 
star-forming galaxies in our survey footprint to a depth of 1-2 Mpc.

\end{abstract}


\section{INTRODUCTION}

	Unbound radial velocities distinguish HVSs from other main sequence stars in
the Galaxy.  In \citet{brown05} we reported the first HVS:  a short-lived \hbox{3
\msun} star traveling with a Galactic rest frame velocity of $700\pm12$ \kms, twice
the Milky Way's escape velocity at the star's distance of $\simeq$100 kpc.  The
observed motion of the star is comparable to the escape velocity from the surface of
the star and is thus difficult to explain with stellar dynamics.  The maximum
ejection velocity from stellar binary disruption mechanisms \citep{blaauw61,
poveda67} is limited to $\sim$300 \kms\ for \hbox{3 \msun} stars \citep{leonard88,
leonard90, leonard91, leonard93, tauris98, portegies00, davies02, gualandris05}.

	There is overwhelming evidence for a $4\times10^6$ M$_{\sun}$ massive black
hole (MBH) in the Galactic center \citep{ghez08, gillessen09}.  The MBH sits in a
vast crowd of stars, including short-lived B stars with orbital periods as short as
15 years \citep{gillessen09b} and pericenter velocities as high as 12,000 \kms\
(=4\% of the speed of light) \citep{ghez05}.  Three-body interactions between stars
and the MBH are inevitable in this environment and will naturally result in the
ejection of unbound ``hypervelocity stars'' \citep{hills88}.  This ejection process
also works for a binary MBH \citep{yu03}.

	A MBH in the Galactic center must produce HVSs, and known HVSs fit the MBH
ejection picture.  The observed unbound velocities, stellar nature, ejection time
distribution, and Galactic latitude and longitude distribution of HVSs all support a
Galactic center origin.  Known HVSs have the spectral types of B stars, the same
spectral type as the stars observed orbiting the central MBH.  In all cases where
photometric variability \citep{fuentes06} or echelle spectroscopy
\citep{przybilla08, przybilla08b, lopezmorales08} is available, HVSs are confirmed
to be short-lived main sequence B stars.
	Even the observed number of HVSs is consistent with the theoretically 
predicted MBH ejection rate \citep{yu03, perets07} and with the number of 
``S-stars'' presently orbiting the MBH \citep{perets09c, bromley12}.  In the Hills 
three-body exchange scenario, S-stars are the former companions of HVSs 
\citep{ginsburg06}, a picture which may \citep{perets08c} or may not 
\citep{madigan11} be supported by the S-stars' eccentricity distribution.  Proper 
motion measurements promise to more directly test the link between HVSs and the MBH 
in the Galactic center \citep[e.g.,][]{brown10a}.

	Here, we present the results of our on-going spectroscopic survey on the
6.5m MMT telescope to find new HVSs.  Our effective survey strategy is to target
stars with the colors of \hbox{$\simeq$3 \msun} stars -- stars bluer than halo blue
horizontal branch (BHB) stars but redder than foreground white dwarfs -- that should
not exist at faint magnitudes unless they were ejected into the outer halo as HVSs.  
In previous papers we reported the discovery of 14 unbound HVSs and a comparable
population of possibly bound HVSs over a fifth of the sky \citep{brown06, brown06b,
brown07a, brown07b, brown09a, brown09b}.  Here we report 5 new unbound HVSs; the
distribution of HVS angular positions on the sky is significantly anisotropic.

	In \S 2 we describe our HVS survey strategy and spectroscopic observations.  
In \S 3 we discuss the distribution of stars in the survey, and identify the 
bound and unbound HVSs.  In \S 4 we discuss the HVSs, and in \S 5 we discuss their 
spatial anisotropy.  We conclude in \S 6.

\section{DATA}

\begin{figure}		
 \includegraphics[width=3.5in]{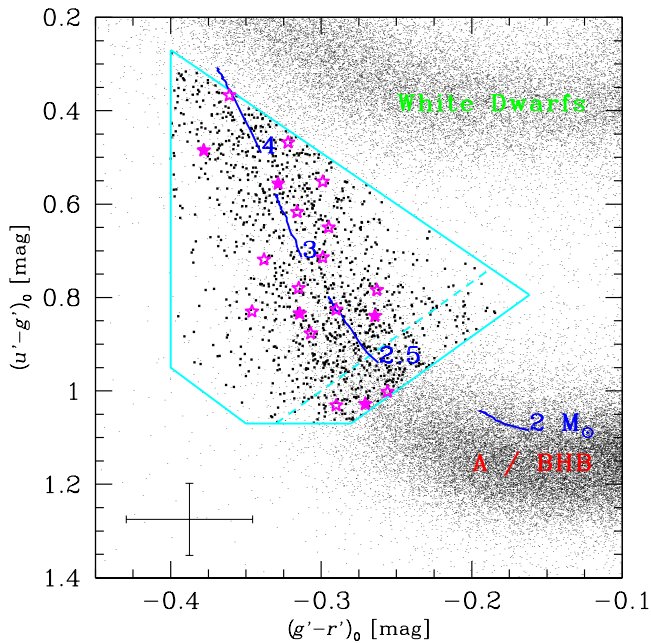}
 \caption{ \label{fig:ugr}
	Color-color diagram showing the target selection for the unified HVS Survey 
(solid cyan line) and the observed candidates (black squares).  The Survey samples 
$17<g<20.25$ to the left of the dashed line and $19<g<20.25$ to the right of the 
dashed line as described in the text.
	Five new HVSs (solid magenta stars) and the fourteen previous HVSs (open 
magenta stars)  scatter around the \citet{girardi04} stellar evolution tracks for 
2.5-4 M$_{\sun}$ main sequence stars (solid blue lines); numbers along the tracks 
indicate stellar mass.  Average SDSS photometric uncertainties for a $g_0=19.5$ star 
are indicated by the errorbars on the lower left.  For reference, we plot the 
underlying distribution of stars in SDSS (black dots) and label the locus of white 
dwarfs and A-type/BHB stars.}
 \end{figure}

\subsection{Target Selection Strategy}

	We select HVS candidates with the magnitudes and colors of \hbox{2.5-4
\msun} stars in the outer halo.  \hbox{2.5-4 \msun} stars are luminous and allow us
to explore a large volume of space.  Selecting these stars maximizes our contrast
with respect to the Milky Way's stellar population, because \hbox{2.5-4 \msun} stars
are bluer than halo BHB stars and redder than disk white dwarfs.  Targeting faint
stars maximizes our efficiency for detecting HVSs because the density of normal
Milky Way stars is very low in the outer halo.  Finally, \hbox{2.5-4 \msun} stars
have main sequence lifetimes of a few $\times10^8$ yr and thus should not exist at
large distances unless they were ejected there.

	We select candidates based on de-reddened point spread function photometry
from the Sloan Digital Sky Survey \citep[SDSS,][]{aihara11}.  Each SDSS data release
changes its photometric calibration, however, which means that each data release
contains candidates unique to that data release plus candidates in common with other
data releases.  Because we observe candidates taken from each subsequent data
release (DR), our HVS Survey is a concatenation of candidates selected from SDSS
DR6, DR7, and DR8.

	We begin with a cut on reddening $E(B-V)<0.1$ to ensure accurate colors.
We exclude the small region of the SDSS between $b<-l/5 + 50\arcdeg$
and $b>l/5-50\arcdeg$ to avoid excessive contamination from Galactic bulge stars.
We also impose $-0.5<(r-i)_0<0$ to eliminate non-stellar objects such as quasars.

	Figure \ref{fig:ugr} illustrates our color selection.  This color selection
unifies the original HVS survey that targets brighter and bluer objects
\citep{brown06, brown06b, brown07a, brown07b} and our more recent HVS survey that
targets fainter and redder objects \citep{brown09a}.  For reference, we plot in
Figure \ref{fig:ugr} the underlying distribution of stars in SDSS and main sequence
tracks for 2 \msun , 2.5 \msun , 3 \msun , and 4 \msun\ solar metallicity stars
\citep{girardi02, girardi04}.  Our color selection region targets the sequence of
2.5 - 4 \msun\ stars bounded by $-0.4 < (g-r)_0 < (-0.43(u-g)_0 +0.18)$ and
$(2.2(g-r)_0 +1.1) < (u-g)_0 < 1.07$.

	Our magnitude selection is $17<g_0<20.25$ in the region $(g-r)_0 <
(-0.43(u-g)_0 +0.13)$.  In the region $(g-r)_0 > (-0.43(u-g)_0 +0.13)$ we restrict
ourselves to fainter magnitudes $18.6 + 10[(g-r)_0+0.3] +
[(u-g)_0-(1.2(g-r)_0+1.25)]< g_0 < 20.25$, or approximately $19<g_0<20.25$.

	Applying these selection criteria to the SDSS DR6, DR7, and DR8 photometric
catalogs results in 874, 1097, and 1424 candidates, respectively.  After removing
objects such as nearby galaxies and bright stars by visual inspection, 
1509 unique candidates remain.  We previously published spectroscopic
identifications and radial velocities for 609 of these candidates (plus a few
hundred others that fall outside the present selection criteria) summarized in
\citet{brown10a}.  SDSS provides spectroscopy for another 63 candidates.  Thus there
remain 837 candidates to be observed; we report observations of 497 of these 
candidates here.

\subsection{Spectroscopic Observations}


	We obtained spectroscopy for 497 candidates at the 6.5m MMT telescope during
observing runs spanning December 2008 to October 2011.  All observations were
obtained with the Blue Channel spectrograph \citep{schmidt89} using the 832 line
mm$^{-1}$ grating in second order and either a 1$\arcsec$ or 1$\farcs$25 slit.  
These settings provide wavelength coverage 3600 \AA\ to 4500 \AA\ and a spectral
resolution of 1 - 1.2 \AA.  All observations were obtained at the parallactic angle
and were paired with comparison lamp exposures.

	Our goal was to obtain modest signal-to-noise ($S/N$) observations adequate
for determining radial velocity.  At $g=19$ mag we typically used a 390 s exposure
to obtain $(S/N)\simeq7$ pix$^{-1}$ in the continuum.  We processed the spectra in
real-time to allow additional observations of interesting candidates.
	We extracted the spectra using IRAF\footnote{IRAF is distributed by the
National Optical Astronomy Observatories, which are operated by the Association of
Universities for Research in Astronomy, Inc., under cooperative agreement with the
National Science Foundation.}
	in the standard way and measured radial velocities using the
cross-correlation package RVSAO \citep{kurtz98}.  The average statistical
uncertainty of the radial velocity measurements is $\pm11$ km s$^{-1}$.

	While we made an effort not to observe stars with existing spectroscopy, 
SDSS re-observed 245 of our stars as part of SEGUE \citep{yanny09}.  We can use
this independent data set to verify our velocity calibration.  The mean difference
between our heliocentric radial velocities and {\it elodiervfinal} from SDSS DR8 is
$2\pm27$ \kms .  The dispersion is consistent with the statistical uncertainties;  
2 \kms\ is our systematic uncertainty.

	Two epochs of spectroscopy also allow us to identify objects with variable
radial velocity.  Although no HVS exhibits a significant change in radial velocity,
there are eight probable white dwarfs that exhibit $>$4-$\sigma$ changes in velocity
between the two epochs.  We expect that these objects are compact binaries, and we
are pursuing follow-up observations to confirm their velocity variability.  To date,
follow-up of low mass \hbox{$<$0.25 \msun} white dwarfs in the HVS Survey has
resulted in the discovery of two dozen merging, double-degenerate binaries with
orbital periods as short as 12 minutes \citep{brown10c, brown11a, brown11b,
brown12a, kilic10, kilic11a, kilic11c, kilic11b, kilic12a}.

\subsection{Survey Completeness and Spectroscopic Identifications}

	With 1169 spectra in hand, the HVS survey is now 90\% complete for DR6- and
DR7-selected candidates and 43\% complete for DR8-selected candidates.  The
remaining candidates are concentrated at faint magnitudes $19.75<g_0<20.25$ but are
spread relatively evenly over the sky.  This distribution of incompleteness is the
result of many poor observing runs that prevented observations of the faintest
objects in the survey.

	Of the 1169 objects: 955 (82\%) are normal stars with late-B spectral types;
192 (16\%) are white dwarfs; 15 (1\%) are quasars at $z\sim2.4$; and 7 (1\%) are B
supergiant stars.  The width and shape of the hydrogen Balmer lines provide a
measure of surface gravity over our range of color (temperature).  We use line
indices described by \citet{brown03} to estimate spectral types, and we perform
stellar atmosphere model fits for the late B-type objects using an upgraded version
of the code described by \citet{allende06}.  The white dwarfs are studied in a
series of papers on low mass white dwarfs that begin with \citet{kilic07, kilic07b}.  
The quasars were reported previously \citep{brown09a}.  We discuss the B supergiants
below, and focus the remainder of this paper on the 955 late B-type stars in the HVS
survey.

\subsection{B Supergiants in the HVS Survey}

	Our spectra reveal seven B supergiants in the $\simeq$10,000 deg$^2$ of sky
sampled by our survey.  All seven B supergiants are associated with known
star-forming galaxies in the Local Group.  The first two B supergiants belong to the
Leo A dwarf.  We used those supergiants plus others to make the first stellar
velocity dispersion measurement of Leo A \citep{brown07c}.

	Four new B supergiants belong to M31 and trace its rotation curve.  The
stars are SDSS J003748.658+395402.57, SDSS J004108.023+401337.08, SDSS
J004828.734+423158.58, and SDSS J004855.598+424629.99.  The four stars have a mean
apparent magnitude of $g_0=19.3$ and thus are located at the distance of M31
\citep{vilardell10} for a typical B9 Ib luminosity of $M_g\simeq-5.0$.  The stars
have angular separations of 1$\arcdeg$ - 2$\arcdeg$ from the center of M31 and
locations consistent with the M31 disk.  The two northern stars have systemic
velocities of +230 \kms\ and +250 \kms\ with respect to M31, and the two southern
stars have systemic velocities of $-126$ \kms\ and $-200$ \kms\ with respect to M31.  
The velocities perfectly match the H{\sc i} rotation curve of M31
\citep{corbelli10}; the B supergiants clearly belong to the outer disk of M31.

	The final B supergiant, SDSS J095951.180+052124.52, lies within 2$\farcm$5 
of the center of Sextans B.  The star's $295\pm4$ \kms\ heliocentric radial velocity 
is identical with the velocity of Sextans B \citep{falco99}.  Thus this B supergiant 
almost certainly belongs to Sextans B, and its existence is further evidence for 
on-going star formation in this Local Group dwarf.  For a distance modulus of 
$(m-M)_0 = 25.56 \pm 0.10$ \citep{sakai97} the $g_0=19.22$ star must have 
$M_g=-6.3$, consistent with a B9 Ia supergiant.  

	With absolute magnitudes of $M_g=-5$ to $-6$, the HVS Survey can detect B9
Ib/a supergiants to a depth of 1-2 Mpc.  The only known star-forming galaxy
undetected in our survey footprint is Pegasus, a dwarf that is known to have an
extremely weak blue plume \citep{massey07}.  M33 also falls in our footprint and was
detected by our color selection, however we excluded M33 stars from the observing
list because of obvious extended emission surrounding the stars.  The absence of B
supergiants elsewhere in the HVS Survey implies that there are no new star-forming
galaxies in our survey footprint to a depth of 1-2 Mpc.

\section{RESULTS}

	We now discuss the properties of our 955 late B-type stars and identify the 
unbound HVSs.  For reference, Galactic escape velocity at the solar circle is 
500-600 \kms\ \citep{smith07} and at 50 kpc it is 300-400 \kms\ \citep{gnedin10}.

\begin{figure}		
 \includegraphics[width=3.5in]{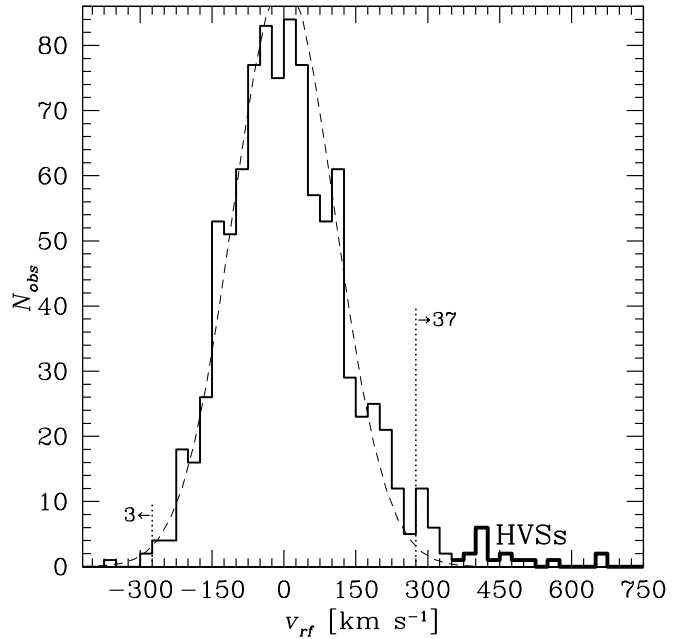}
 \caption{ \label{fig:velh}
	Minimum Galactic rest-frame velocity $v_{rf}$ distribution for the 955 late
B-type stars in the HVS survey.  The best-fit Gaussian (dashed line) has dispersion
$107$ km s$^{-1}$, excluding the 39 stars with $|v_{rf}|>275$ \kms.  The significant
absence of $v_{rf}<-275$ \kms\ stars demonstrates that the positive velocity
outliers are short-lived.  Stars with $v_{rf}>400$ \kms\ are unbound.}
 \end{figure}

\subsection{Radial Velocity Distribution}

	Figure \ref{fig:velh} plots the distribution of line-of-sight velocities
corrected to the Galactic rest-frame $v_{rf}$ for the 955 late B-type stars in the
HVS Survey.  We calculate rest frame velocities using the local standard of rest
from \citet{schonrich10} and a 250 \kms\ circular rotation velocity \citep{reid09,
mcmillan10}:
	\begin{equation} v_{rf} = v_{helio} + 11.1\cos{l}\cos{b} +  
262.24\sin{l}\cos{b} + 7.25\sin{b}. \label{eqn:vlsrh} \end{equation}
	This definition of $v_{rf}$ changes our previous estimates of Galactic
rest-frame velocity by up to 30 \kms.  We note that $v_{rf}$ represents the radial
component of velocity in the Galactic rest frame, and is thus a lower limit to the
true velocity of any HVS.

	The 915 survey stars with $|v_{rf}|<275$ \kms\ have a $0\pm4$ \kms\ mean and
a $107\pm5$ \kms\ dispersion, consistent with a stellar halo population.  The
velocity distribution, however, is not exactly Gaussian (Figure \ref{fig:velh}).  
\citet{king12} show that 10-17\% of the HVS Survey stars may belong to the
Sagittarius stream, which explains the excess of stars in the range $-50$ to $-150$
\kms\ and at +100 \kms.

	Stars with $v_{rf}>+400$ \kms\ are unbound and probable HVSs.  We observe no
stars moving towards us with $v_{rf}<-400$ \kms, consistent with the picture that
stars with $v_{rf}>+400$ \kms\ are ejected from the Milky Way.  The most negative
velocity star in our sample has $v_{rf}=-359\pm10$ \kms, consistent with
\citet{kenyon08}'s estimate of escape velocity at 50 kpc.  If the unbound stars are
HVSs, then HVS ejection models show that we must also find a comparable number of
bound HVSs in our survey \citep{bromley06, kenyon08}.

	Indeed, we observe a significant excess of stars around +300 \kms\ that are 
possibly bound HVSs.  As an estimate of significance, there is less than a $10^{-5}$ 
probability of randomly drawing 18 stars with $275<v_{rf}<325$ \kms\ from the tail 
of a Gaussian distribution with the observed parameters.  Thus the excess of stars 
around +300 \kms\ are probably not halo stars.  Because the stars around +300 \kms\ 
are bound to the Milky Way, the absence of a comparable number of stars at $-300$ 
\kms\ demonstrates that the bound stars must have main-sequence lifetimes less than 
their $\sim$1 Gyr orbital turn-around times \citep{brown07b, kollmeier07, yu07}.  
We simply do not see stars falling back onto the Galaxy with similar velocities.  
Given the observed colors and spectra, we conclude that the stars around +300 \kms\ 
are main-sequence \hbox{$\simeq$3 \msun} stars ejected into the outer halo.

\begin{figure}		
 \includegraphics[width=3.5in]{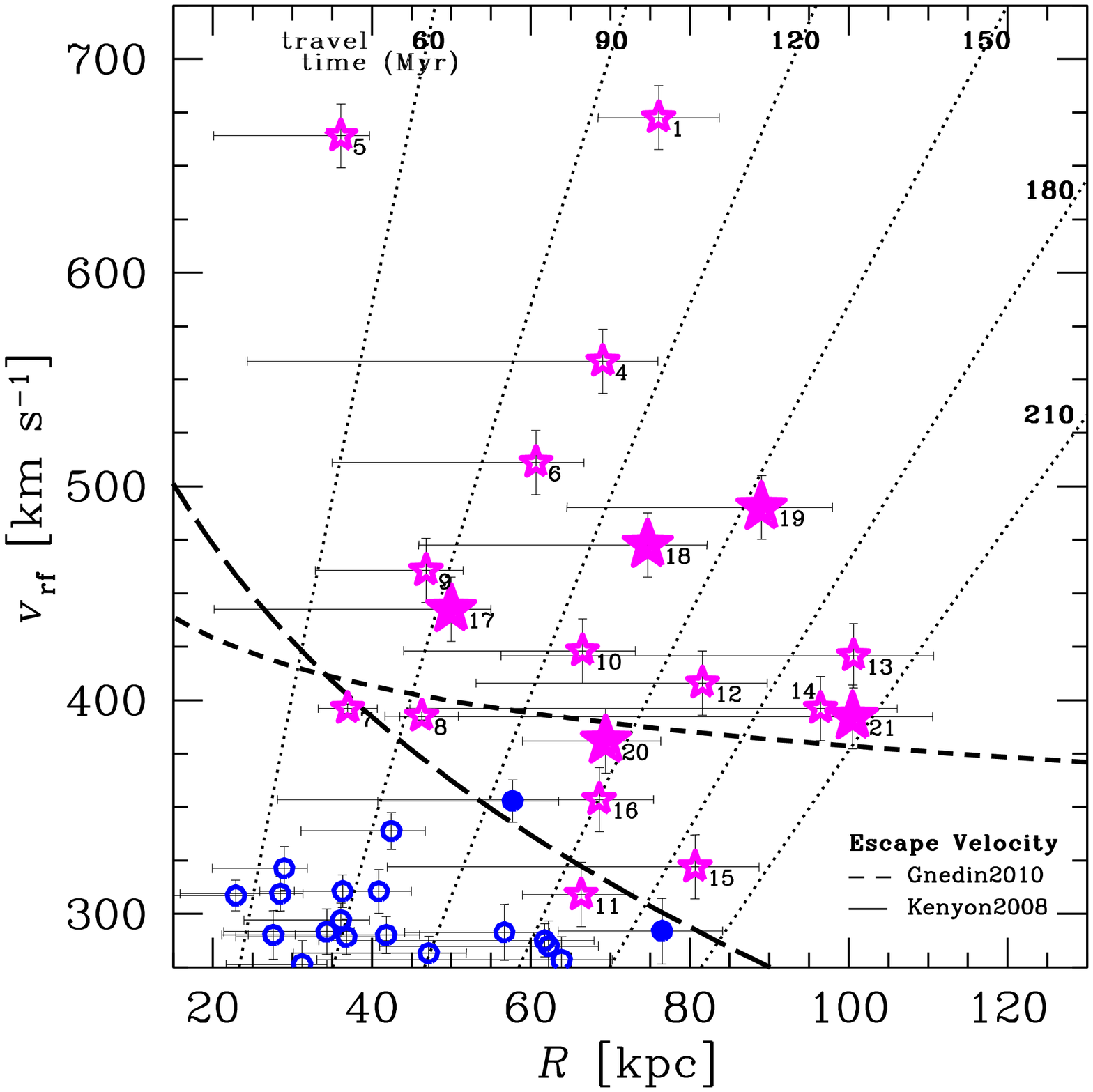}
 \caption{ \label{fig:travel}
	Minimum rest-frame velocity vs.\ Galactocentric distance $R$ for all 37 
stars with $v_{rf}>+275$ \kms.  Distances are estimated using \citet{girardi04} main 
sequence star tracks.  Errorbars show the span of physically possible distance given 
by evolved BHB tracks \citep{dotter08}. Five new HVSs (solid magenta stars) have 
velocities and distances exceeding the \citet{gnedin10} escape velocity model 
(dashed line).  Other possible HVSs (solid blue dots) have velocities and distances 
near the \citet{kenyon08} escape velocity model (long dashed line).  Previously 
identified HVSs are marked with open magenta stars.
	Isochrones of travel time from the Galactic center (dotted lines) are
calculated assuming the observed minimum rest frame velocity $v_{rf}$ is the full 
space motion of the stars.  }
\end{figure}

\subsection{Unbound Stars}

	We require distance estimates to determine whether a high velocity star is 
unbound.  The observations demonstrate that high velocity stars are short-lived main 
sequence stars.  We estimate luminosities by comparing observed colors to 
\citet{girardi02, girardi04} stellar evolution tracks (see Figure \ref{fig:ugr}) for 
solar abundance main sequence stars.  We use both $(u-g)_0$ and $(g-r)_0$ colors to 
estimate luminosity as described in \citet{brown10a}.  We estimate that the 
precision of our luminosities is $\pm$0.23 mag based on propagating color errors 
through the stellar evolution tracks with a Monte Carlo technique.  We then 
calculate heliocentric distances using de-reddened apparent magnitudes, and convert 
to Galactocentric distances $R$ assuming that the Sun is 8 kpc from the Galactic 
center.

	Systematics are the dominant source of uncertainty in our distance 
estimates.  The \citet{girardi02, girardi04} absolute magnitudes are $0.24\pm0.33$ 
mag fainter than the \citet{schaller92} absolute magnitudes that we used in earlier 
HVS survey papers.  Thus our new luminosity estimates shift known HVSs to smaller 
distances.  HVS7 suffers the biggest change, going from 60 kpc to 37 kpc; echelle 
spectroscopy, however, establishes that HVS7 is a \hbox{3.7 \msun} Bp star with 
$M_V\simeq-1.05$ \citep{przybilla08b} and thus $\simeq$60 kpc distant.  This example 
highlights the uncertainty inherent in stellar evolution tracks.  Stellar nature is 
another contributor to systematic uncertainty.  We estimate a physical lower 
distance limit for our stars by assuming that they are evolved BHB stars.  
\citet{dotter07, dotter08} tracks show that hot, metal-poor BHB stars are 
$1.33\pm0.69$ mag fainter than a main sequence star of the same temperature in our 
survey.  We indicate BHB distances with the horizontal error bars in Figure 
\ref{fig:travel}.

\begin{deluxetable*}{lccccccccc}           
\tabletypesize{\footnotesize}
\tablewidth{0pt}
\tablecaption{HVS Survey Stars with $v_{rf}>+275$ \kms\label{tab:hvs}}
\tablecolumns{10}
\tablehead{
  \colhead{ID} & \colhead{$v_{\sun}$} & \colhead{$v_{rf}$} & 
  \colhead{$g_0$} & \colhead{$M_g$} & \colhead{$R_{GC}$} & 
  \colhead{$l$} & \colhead{$b$} & \colhead{Catalog} & \colhead{Ref} \\
  \colhead{} & \colhead{(\kms )} & \colhead{(\kms )} &
  \colhead{(mag)} & \colhead{(mag)} & \colhead{(kpc)} &
  \colhead{(deg)} & \colhead{(deg)} & \colhead{} & \colhead{}
}
	\startdata
\cutinhead{HVSs}
 HVS1  & 840 & 673  & 19.69 & $+0.41$ &  76  & 227.33 & $+31.33$ & SDSS J090744.99$+$024506.88 & 1 \\
 HVS2  & 708 & 718  & 19.05 & $+2.6 $ &  26  & 175.99 & $+47.05$ &                    US708    & 2 \\
 HVS3  & 723 & 520  & 16.20 & $-2.7 $ &  62  & 263.04 & $-40.91$ &               HE0437$-$5439 & 3 \\
 HVS4  & 611 & 559  & 18.31 & $-0.66$ &  69  & 194.76 & $+42.56$ & SDSS J091301.01$+$305119.83 & 4 \\
 HVS5  & 553 & 664  & 17.56 & $+0.04$ &  36  & 146.23 & $+38.70$ & SDSS J091759.47$+$672238.35 & 4 \\
 HVS6  & 626 & 511  & 18.97 & $+0.10$ &  61  & 243.12 & $+59.56$ & SDSS J110557.45$+$093439.47 & 5 \\
 HVS7  & 529 & 396  & 17.64 & $-0.19$ &  37  & 263.83 & $+57.95$ & SDSS J113312.12$+$010824.87 & 5 \\
 HVS8  & 489 & 393  & 17.94 & $+0.01$ &  46  & 211.70 & $+46.33$ & SDSS J094214.03$+$200322.07 & 6 \\
 HVS9  & 628 & 461  & 18.64 & $+0.34$ &  47  & 244.63 & $+44.38$ & SDSS J102137.08$-$005234.77 & 6 \\
HVS10  & 478 & 423  & 19.22 & $+0.21$ &  66  & 249.93 & $+75.72$ & SDSS J120337.85$+$180250.35 & 6 \\
HVS12  & 552 & 408  & 19.61 & $+0.13$ &  82  & 247.11 & $+52.46$ & SDSS J105009.59$+$031550.67 & 7 \\
HVS13  & 575 & 421  & 20.02 & $+0.09$ & 101  & 251.65 & $+50.64$ & SDSS J105248.30$-$000133.94 & 7 \\
HVS14  & 532 & 396  & 19.72 & $-0.25$ &  96  & 241.78 & $+53.20$ & SDSS J104401.75$+$061139.02 & 7 \\
HVS17  & 246 & 442  & 17.43 & $-0.74$ &  50  &  73.52 & $+41.16$ & SDSS J164156.39$+$472346.12 &   \\
HVS18  & 251 & 473  & 19.30 & $+0.05$ &  75  & 103.64 & $-26.77$ & SDSS J232904.94$+$330011.47 &   \\
HVS19  & 597 & 490  & 20.06 & $+0.35$ &  89  & 256.05 & $+63.74$ & SDSS J113517.75$+$080201.49 &   \\
HVS20  & 504 & 381  & 19.81 & $+0.54$ &  69  & 262.56 & $+60.39$ & SDSS J113637.13$+$033106.84 &   \\
HVS21  & 355 & 392  & 19.73 & $-0.23$ & 100  & 165.26 & $+56.11$ & SDSS J103418.25$+$481134.57 &   \\
\cutinhead{Possible HVSs}
HVS11  & 477 & 307  & 19.58 & $+0.68$ &  64  & 238.77 & $+40.63$ & SDSS J095906.47$+$000853.41 & 7 \\
       & 222 & 292  & 19.83 & $+0.53$ &  77  & 155.50 & $+49.46$ & SDSS J101359.79$+$563111.65 & 7 \\
       & 496 & 353  & 18.86 & $+0.28$ &  54  & 256.27 & $+54.55$ & SDSS J111136.44$+$005856.44 &   \\
HVS15  & 463 & 322  & 19.15 & $-0.29$ &  78  & 266.51 & $+55.92$ & SDSS J113341.09$-$012114.25 & 7 \\
HVS16  & 434 & 345  & 19.33 & $-0.03$ &  74  & 285.86 & $+67.38$ & SDSS J122523.40$+$052233.84 & 7 \\
\cutinhead{Possible Bound HVSs}
       & 150 & 321  & 17.14 & $+0.29$ &  27  & 115.82 & $-40.58$ & SDSS J002810.33$+$215809.66 &   \\
       & 136 & 311  & 17.77 & $+0.03$ &  40  & 125.05 & $-31.26$ & SDSS J005956.06$+$313439.29 &   \\
       & 108 & 276  & 17.19 & $+0.18$ &  30  & 127.53 & $-31.42$ & SDSS J010948.30$+$311727.66 &   \\
       & 361 & 288  & 18.39 & $-0.25$ &  61  & 196.07 & $+23.21$ & SDSS J074950.24$+$243841.16 & 7 \\
       & 229 & 297  & 17.28 & $+0.06$ &  34  & 160.45 & $+34.20$ & SDSS J081828.07$+$570922.07 & 7 \\
       & 306 & 282  & 18.08 & $+0.14$ &  45  & 186.30 & $+42.16$ & SDSS J090710.07$+$365957.54 & 7 \\
       & 504 & 339  & 18.48 & $+0.39$ &  44  & 247.97 & $+46.42$ & SDSS J103357.26$-$011507.35 &   \\
       & 448 & 285  & 19.23 & $+0.14$ &  68  & 250.71 & $+47.87$ & SDSS J104318.29$-$013502.51 &   \\
       & 482 & 309  & 17.38 & $+0.08$ &  30  & 269.75 & $+47.30$ & SDSS J112255.77$-$094734.92 &   \\
       & 140 & 291  & 19.13 & $+0.47$ &  58  & 130.08 & $+40.59$ & SDSS J112359.47$+$751807.73 &   \\
       & 424 & 290  & 18.13 & $-0.07$ &  44  & 274.88 & $+57.45$ & SDSS J115245.91$-$021116.21 & 7 \\
       & 228 & 309  & 17.49 & $+0.06$ &  31  &  65.34 & $+72.37$ & SDSS J140432.38$+$352258.41 & 7 \\
       & 284 & 289  & 18.40 & $+0.22$ &  40  & 357.16 & $+63.62$ & SDSS J141723.34$+$101245.74 & 7 \\
       & 206 & 278  & 18.88 & $-0.12$ &  58  &  18.68 & $+44.85$ & SDSS J154806.92$+$093423.93 &   \\
       &  62 & 292  & 17.51 & $+0.12$ &  29  &  75.71 & $+28.06$ & SDSS J180050.86$+$482424.63 &   \\
       & 133 & 290  & 17.35 & $+0.41$ &  25  &  85.47 & $-51.67$ & SDSS J232229.47$+$043651.45 &   
	\enddata
\tablerefs{ (1) \citet{brown05}; (2) \citet{hirsch05}; (3) \citet{edelmann05};
(4) \citet{brown06}; (5) \citet{brown06b}; (6) \citet{brown07b}; (7) \citet{brown09a} }
 \end{deluxetable*}

	Figure \ref{fig:travel} plots the resulting distribution of minimum Galactic
rest frame velocity $v_{rf}$ versus Galactocentric distance $R$ for the 37 stars
with $v_{rf}>+275$ \kms.  Identifying the unbound stars requires a potential model
for the Galaxy.  Because the Galactic potential is poorly constrained at large
distances, we consider two recent models.

	\citet{kenyon08} construct a spherically symmetric potential to fit observed
Milky Way mass measurements from 5 pc to 10$^5$ pc.  We estimate escape velocity
from this model by dropping a test particle from rest at 250 kpc and calculating its
infall velocity.  This calculation yields an escape velocity of 360 \kms\ at 50 kpc
(long-dashed line in Figure \ref{fig:travel}).  \citet{gnedin10} measure the mass
profile of the Milky Way using the velocity dispersion of the (bound) HVS Survey
stars.  Assuming that the escape velocity is twice the circular velocity, we
estimate an escape velocity of 400 \kms\ at 50 kpc (dashed line in Figure
\ref{fig:travel}).  The escape velocity profiles differ in part because the
\citet{gnedin10} dark matter halo is $1.6\times10^{12}$ \msun\ while the
\citet{kenyon08} dark matter halo is $1\times10^{12}$ \msun.  The difference in
escape velocity profiles illustrates the present uncertainties.  Because a larger
halo mass is preferred by recent observations \citep[e.g.,][]{przybilla10}, we use
the \citet{gnedin10} model to identify new unbound stars.

	We identify five new unbound HVSs with radial velocities and distances
equaling or exceeding the \citet{gnedin10} escape velocity model, and one new
possible HVS exceeding the \citet{kenyon08} escape velocity model.  Previously we
identified 14 unbound HVSs on the basis of the \citet{kenyon08} model, however our
new $v_{rf}$ and distance estimates place HVS11 slightly below the threshold of the
\citet{kenyon08} model.  Another two stars, HVS15 and HVS16, are above the threshold
of the \citet{kenyon08} model but well below the threshold of the \citet{gnedin10}
model.  Therefore we re-classify these three objects as ``possible HVSs.'' HVS7 and
HVS8, on the other hand, are demonstrated \hbox{$\simeq$4 \msun} main sequence B
stars at $R\simeq$50 kpc \citep{lopezmorales08, przybilla08b} and thus are almost
certainly HVSs.

	Table \ref{tab:hvs} summarizes the properties of all 37 HVS Survey stars
with $v_{rf}>+275$ \kms.  Columns include HVS number, heliocentric radial velocity
$v_{\sun}$, minimum Galactic rest-frame velocity $v_{rf}$ (not a full space
velocity), de-reddened SDSS $g$-band magnitude, absolute magnitude $M_g$ from
\citet{girardi02, girardi04} main sequence tracks, Galactocentric distance $R$,
Galactic coordinates $(l,b)$, and catalog identification.  There are 16 unbound
HVSs, 5 possible HVSs, and 16 possibly bound HVSs.

	The velocity and distance estimates allow us to estimate ejection times.  
The dotted lines in Figure \ref{fig:travel} indicate travel times from the Galactic
center under the assumption that $v_{rf}$ is the 3-dimensional velocity of our
stars, a conservative assumption even though the orbits should be nearly radial
\citep{gnedin05}.  We find that HVSs are not clustered around a common ejection
time, but have a range of ejection times spread continuously between 60 and 210 Myr.  
In other words, the observations do not support a large burst of HVSs that one might
expect for binary MBH in-spiral or dwarf galaxy tidal disruption models.  
Rather, the observations support an on-going ejection process, consistent with the
Hills three-body ejection model.

\section{FIVE NEW HYPERVELOCITY STARS}

	We discuss the details of the five new unbound HVSs discovered in this
portion of the survey.  All of the HVSs except for HVS17 have multiple epochs of
observation.  None of the HVSs exhibit significant radial velocity variations, which
indicates that the HVSs are unlikely to be compact binaries.

	SDSS J164156.391+472346.12, hereafter HVS17, has a $+246\pm9$ \kms\
heliocentric radial velocity and a minimum velocity of $v_{rf}=+443$ in the Galactic
frame.  HVS17 is the hottest HVS in our survey with a B6 spectral type.  HVS17 is
also the brightest HVS in our survey, $g=17.499\pm0.015$ mag, opening the
possibility of follow-up echelle spectroscopy.  A solar metallicity \hbox{4 \msun}
star with the temperature of HVS17 has $M_g=-0.74$ \citep{girardi02, girardi04} and
a distance of $R=50$ kpc.

	SDSS J232904.947+330011.47, hereafter HVS18, has a $+251\pm10$ \kms\
heliocentric radial velocity and a minimum velocity of $v_{rf}=+473$ \kms\ in the
Galactic frame.  At $b=-26.8\arcdeg$, HVS18 is the lowest latitude HVS in our
survey. HVS18 is also our first HVS discovery in the southern Galactic hemisphere.  
Unlike HVS3, the \hbox{9 \msun} HVS near the LMC \citep{edelmann05, przybilla08},
HVS18 is 17\arcdeg\ from M31.  Any association with M31 or its satellites is
impossible.  A solar metallicity \hbox{3 \msun} star with the temperature of HVS18
has $M_g=0.05$, thus HVS18 has a distance of $R=75$ kpc from the Milky Way and a
distance of 700 kpc from M31.

	SDSS J113517.759+080201.49, hereafter HVS19, has a $+597\pm15$ \kms\
heliocentric radial velocity and a minimum velocity of $v_{rf}=+490$ \kms\ in the
Galactic frame.  HVS19 is one of the cooler HVSs in our survey with an A0 spectral
type.  A solar metallicity \hbox{$\simeq$2.5 \msun} star with the temperature of
HVS19 has $M_g=0.35$ which locates it at $R=90$ kpc.  Like many of the other HVSs,
HVS19 is in the constellation of Leo.

	SDSS J113637.135+033106.84, hereafter HVS20, has a $504\pm12$ \kms\
heliocentric radial velocity and a minimum velocity of $v_{rf}=+381$ \kms\ in the
Galactic frame.  HVS20 sits on the escape velocity threshold of \citet{gnedin10} but
is well above the threshold of \citet{kenyon08}.  HVS20 is similar to HVS19, only
4$\fdg$5 away; it is yet another HVS in the constellation Leo.  HVS20 is also one of
the coolest HVSs in our survey with an A1 spectral type.  With an estimated
luminosity of $M_g=0.54$, HVS20 sits at $R=80$ kpc and so shares a similar ejection
time as HVS19.  The other eight HVSs around Leo have ejection times that differ by
up to 125 Myr, however, implying that the HVSs in the direction of Leo do not all
come from a single ejection event.

	SDSS J103418.254+481134.57, hereafter HVS21, has a $+355\pm10$ \kms\
heliocentric radial velocity and a minimum velocity of $v_{rf}=+392$ \kms\ in the
Galactic frame.  HVS21 has a B7 spectral type and, for a luminosity of $M_g=-0.23$,
is located at $R=100$ kpc.  Like HVS20, HVS21 sits on escape velocity threshold of
\citet{gnedin10} but is well above the threshold of \citet{kenyon08}. HVS21 is in
Ursa Major about 11\arcdeg\ from HVS2 \citep{hirsch05}.

\begin{figure*}		
 \includegraphics[width=3.5in]{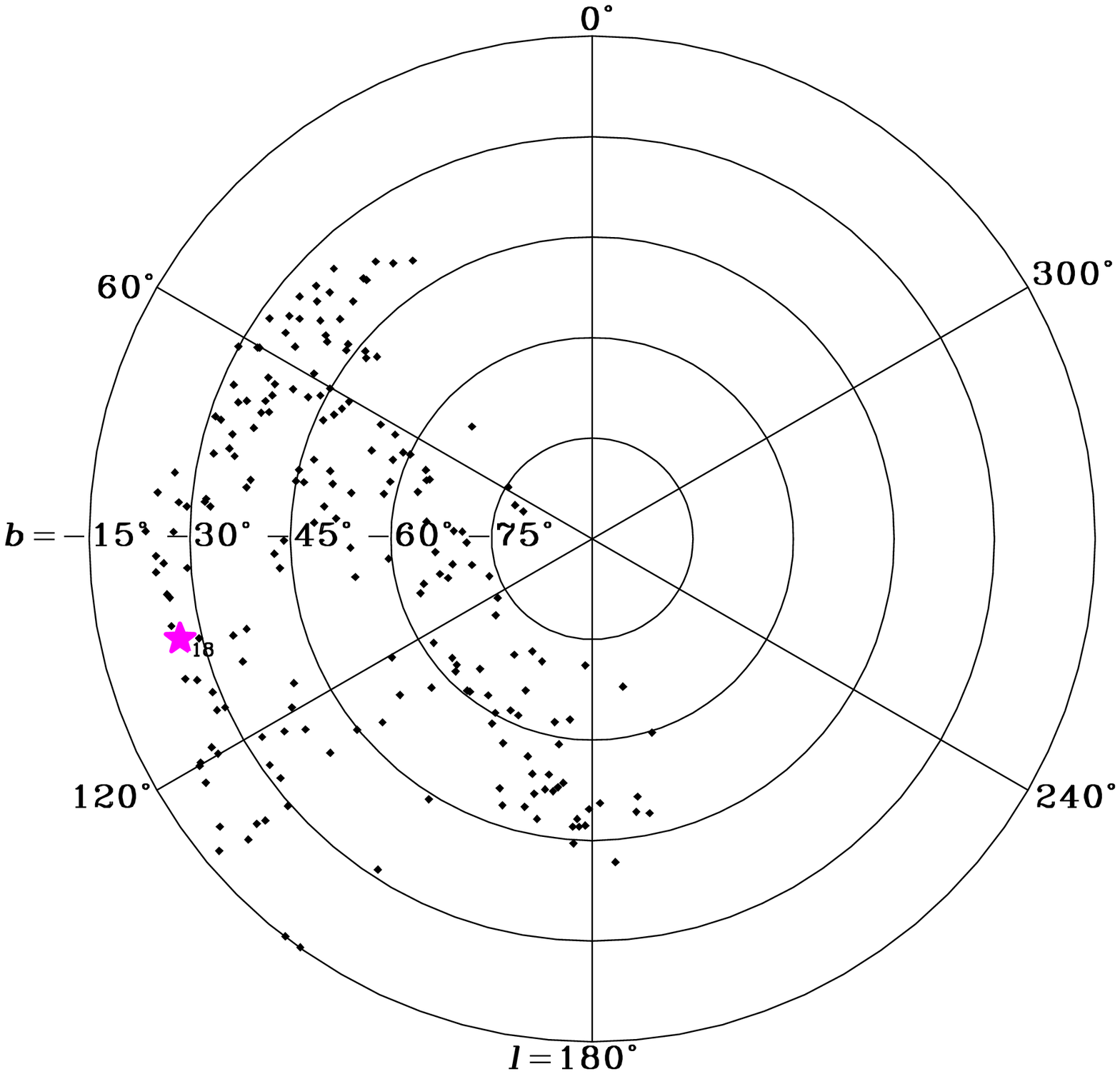}\includegraphics[width=3.5in]{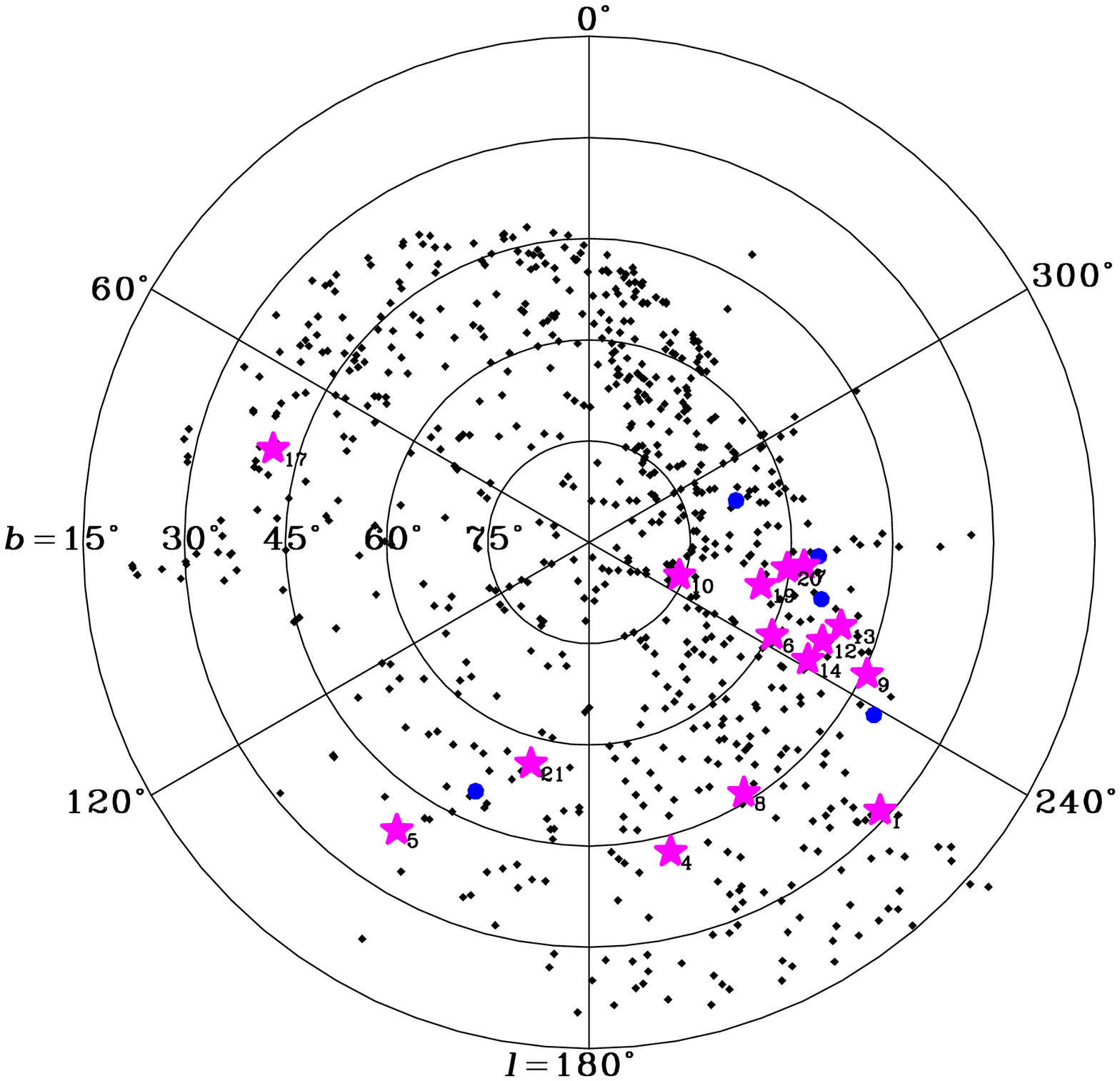}
 \caption{ \label{fig:polar}
	Polar projections, in Galactic coordinates, showing the 16 unbound HVSs 
(magenta stars), the 5 possible HVSs (blue circles) and 934 other stars (black dots) 
in the HVS survey over the southern (left) and northern (right) Galactic caps.}
 \end{figure*}

\begin{figure}		
 \includegraphics[width=3.25in]{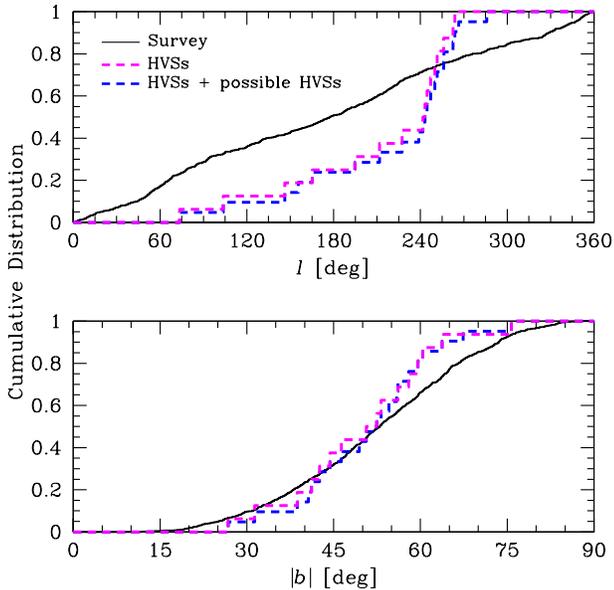}
 \caption{ \label{fig:gll}
	Cumulative distributions of Galactic $l$ and $b$ of the 16 unbound HVSs 
(dashed magenta line), 21 HVSs and possible HVSs (dashed blue line), and the other 
stars (solid line) in the HVS survey.}
 \end{figure}

\section{SPATIAL ANISOTROPY}

	The spatial distribution of HVSs on the sky is intriguing because it is 
almost certainly linked to their origin.  In principle, the MBH in the Galactic 
center can eject HVSs in all directions.  \citet{brown09b} demonstrate that unbound 
HVSs have a significantly anisotropic distribution on the sky.  Lower velocity bound 
HVSs have a more isotropic distribution.  The new HVSs reported here support these 
trends.

	Figure \ref{fig:polar} plots the angular positions of all of the survey
stars.  The overall distribution of positions reflects the SDSS imaging footprint.  
HVSs are marked with solid magenta stars and possible HVSs are marked with solid
blue dots.  The stars have a similar distribution and so we consider the combined
set of HVSs in an attempt to mitigate small number statistics.  Because of the
complex SDSS footprint and because of the HVS Survey incompleteness, we quantify the
spatial anisotropy by comparing the distribution of HVSs against the distribution of
observed stars from which they are drawn.  Although this comparison is not perfect
-- the overdensity of stars arcing to the right of the North Galactic pole in Figure
\ref{fig:polar} is the Sgr stream -- this approach provides a fair comparison of
positions and angular separations in the HVS survey.

	Figure \ref{fig:gll} plots the cumulative Galactic longitude and latitude 
distributions of our survey stars and HVSs.  Kolmogorov--Smirnov (K-S) tests find 
0.013 and 0.415 likelihoods that the combined set of HVSs are drawn from the same 
longitude and latitude distributions, respectively, as the survey stars.  The 
longitude and latitude tests are not completely independent because of the complex 
boundaries of our survey region.  The KS tests establish, however, that the HVS 
spatial anisotropy is primarily in Galactic longitude and not in Galactic latitude.

	Next we consider the distribution of angular separations.  Seven of the HVSs 
(33\%) are in the constellation Leo, thus we expect that the HVSs have an excess of 
small angular separations and an absence of large angular separations compared to 
the survey stars.  Calculating the angular separations for all unique pairs of 
stars, a K-S test finds a $3\times10^{-18}$ likelihood that the HVSs are drawn from 
the same distribution of angular separations as the survey stars.  We establish the 
significance of this likelihood by taking 10,000 random draws of 21 unique stars 
from the survey: a likelihood below $3\times10^{-18}$ occurs at random 0.3\% of the 
time in our survey.  The actual significance of the anisotropy depends on the method 
of calculation; our approach indicates a formal significance of 3-$\sigma$.

	A number of models propose to explain the HVS spatial anisotropy.  
\citet{abadi09} propose a tidal debris origin for HVSs \citep[see also][]{piffl11}.
However, this model is difficult to reconcile with the observations because 1) known
HVSs have ejection times that differ by up to 150 Myr, 2) no other unbound tidal
debris is observed in the same region of sky \citep{kollmeier09, kollmeier10}, and
3) no dwarf galaxy in the Local Group has comparable velocity.  An intermediate mass
black hole in-spiral event in the Galactic center may produce a ring of HVSs around
the sky \citep{gualandris05, levin06, sesana06, baumgardt06} but this model is also
at variance with the observations.  Theoretically predicted black hole in-spirals
occur on timescales 10--100 times shorter than the observed span of HVS ejection
times, thus multiple black hole in-spiral events with common alignment are necessary
to account for the observed spatial distribution.

	Runaway stars, stars born in the disk and ejected by binary disruptions, are
another possible source of apparent HVSs \citep[e.g.][]{gvaramadze09}.  Surveys of
stars near the Galactic disk identify unbound stars ejected from the disk
\citep{heber08, przybilla08c, tillich09, irrgang10} as expected in runaway star
mechanisms \citep{bromley09, gvaramadze09b}.  Stars ejected from a rotating disk
have a distinct spatial distribution:  the fastest stars are those ejected in the
direction of Galactic rotation, thus $\simeq$90\% of $>$400 \kms\ runaways should be
located $|b|<45\arcdeg$ \citep{bromley09}.  Moreover, $>$400 \kms\ runaways at
$R=50$ kpc should be found over all Galactic longitudes.  Thus the expected
distribution of runaway latitudes and longitudes is inconsistent with the observed
distribution of HVSs.

	The ejection {\it rate} of disk runaways is also inconsistent with the
observed number of HVSs.  The physical size of stars places a speed limit on stellar
binary disruption mechanisms:  a \hbox{3 \msun} runaway with $v_{rf}>+400$ \kms\ at
$R=50$ kpc requires the disruption of a compact binary containing a massive star
within its lifetime while avoiding a merger.  An optimistic ejection rate for
\hbox{3 \msun} runaways with $>$400 \kms\ at $R=50$ kpc is about 100 to 100,000
times smaller than the expected rate of HVSs ejected by the MBH \citep{brown09a,
perets12}.  Given that the HVSs are not clustered at low latitudes in our
survey, we conclude that runaways are unlikely contaminants to the HVS sample.

	The anisotropic distribution of stars in the Galactic center provides
another model for explaining the anisotropy of HVSs.  If the central MBH is the
origin of HVSs, and the stars orbiting near the MBH are concentrated in one or two
disks \citep{levin03, lu09, bartko09}, then stars probably interact with the MBH in
preferred planes.  \citet{lu10} and \citet{zhang10} demonstrate that the observed
distribution of HVSs is consistent with two thin disk planes aligned with the
present stellar disks in the Galactic center.  Observed HVSs were ejected up to 200
Myr ago, however.  Although the Galactic center clearly contains anisotropic
structure -- including young stellar clusters like the Arches and Quintuplet
clusters and gaseous disks, arms, and bars that may form young stars -- there is no
model that explains how Galactic center structure maintains a fixed orientation for
200 Myr.

	A final model is an anisotropic Galactic potential.  HVSs are unique test
particles because they start in the Galactic center and travel to infinity.  
The gravitational potential maps initial HVS ejection velocities to the velocities
that we observe in the outer halo today.  Because many HVSs are marginally unbound,
a non-spherical potential can naturally explain why HVSs are found in preferred
directions on the sky:  stars ejected along the long axis of the potential would
have the fastest observed velocities; stars ejected along the minor axis of the
potential are decelerated more and so fewer will be observed as HVSs.  
\citet{kenyon08} show that the important quantity is the scale length of potential
in the central 10-100 pc, the region where HVSs experience the most deceleration.  
The anisotropic potential model can be tested with the all-sky distribution of HVSs.
If there is rotation around the long axis of the potential, stars ejected along the
long axis should be located in two clumps on opposite points in the sky.  If there
is rotation around the short axis, stars ejected along the long axis should appear
in ring of HVSs around the sky.  There are currently no good constraints on the
shape and orientation of the Galactic potential, but perhaps the HVSs will reveal
it.

\section{CONCLUSIONS}

	We report new results from our targeted HVS Survey, a spectroscopic survey
of stars with \hbox{$\simeq$3 \msun} colors that should not exist at faint
magnitudes unless they were ejected into the outer halo as HVSs.  Our survey is 90\%
complete for SDSS DR6- and DR7-selected candidates and 43\% complete for
DR8-selected candidates.  Spectroscopy reveals that most of the candidates are
normal late B-type stars in the halo; other objects include low mass white dwarf
binaries and B supergiants in known star-forming galaxies.  The absence of B
supergiants elsewhere in the HVS Survey implies that there are no new star-forming
galaxies to a depth of 1-2 Mpc within our survey footprint.

	The velocity distribution of the 955 late B-type stars in the survey is
remarkable: 37 stars have $v_{rf}>+275$ km s$^{-1}$ and 3 stars have $v_{rf}<-275$
km s$^{-1}$.  The observational signature of a HVS is its unbound velocity, which we
determine by comparing observed radial velocities and distances to Galactic
potential models.  There are 5 new unbound HVSs on the basis of the \citet{gnedin10}
escape velocity model.  The HVS Survey thus identifies 16 unbound HVSs, 5 possible
HVSs, and 16 possibly bound HVSs.

	A Galactic center origin for the HVSs is supported by their unbound
velocities, the observed number of unbound stars, their stellar nature, their
ejection time distribution, and their Galactic latitude and longitude distribution.  
Other proposed origins for the unbound HVSs, including runaway ejections from the
disk or dwarf galaxy tidal debris, cannot easily be reconciled with the
observations.  Although not all unbound stars are necessarily HVSs, the MBH ejection
origin provides the best explanation for the observed \hbox{$\simeq$3 \msun} unbound
HVSs in our survey.

	HVSs are important because their properties are tied to the nature and 
environment of the MBH that ejects them \citep{portegies06, merritt06, demarque07, 
ginsburg07, gualandris07, hansen07, lu07, sesana07, sesana07b, sesana08, sesana09, 
oleary08, perets08a, perets09c, perets09a, sherwin08, svensson08, lockmann08, 
lockmann08b, lockmann09, chen09, gualandris09, hopman09, madigan09, sari10, 
antonini10, antonini11, baruteau11, ginsburg12}.  HVS are also important probes of 
the dark matter potential through which they move \citep{gnedin05, yu07, wu08, 
kenyon08, perets09b}.

	An intriguing result of the HVS Survey is the spatial anisotropy of unbound
HVSs on the sky.  There is as yet no compelling explanation for the anisotropy.  We
speculate that an anisotropic central potential may explain the observations.  In
the future, measuring the full space trajectories of the HVSs promises to elucidate
the origin of HVSs \citep{brown10b} and better constrain the shape and orientation
of the Galactic potential.  Further progress requires measuring the spatial
distribution of HVSs over the entire sky.  We look forward to using the Skymapper
Survey \citep{keller07} to identify HVS candidates in coming years.

\acknowledgements

	We thank M.\ Alegria, A.\ Milone, and J. McAfee for their assistance with 
observations obtained at the MMT Observatory, a joint facility of the Smithsonian 
Institution and the University of Arizona.  This project makes use of data products 
from the Sloan Digital Sky Survey, which is managed by the Astrophysical Research 
Consortium for the Participating Institutions.  This research makes use of NASA's 
Astrophysics Data System Bibliographic Services.  This work was supported by the 
Smithsonian Institution.

{\it Facilities:} \facility{MMT (Blue Channel Spectrograph)}

\clearpage


\appendix				
\section{DATA TABLE}

	Table \ref{tab:dat} presents the sample of 955 late B-type stars in the HVS
Survey studied here.  For completeness, we provide both the 375 newly observed
objects and the 580 previously reported objects.  Columns include J2000 coordinates,
SDSS $g$-band apparent magnitude, our heliocentric velocity, velocity in the
Galactic rest frame (Equation \ref{eqn:vlsrh}), and Galactic longitude and latitude.

\begin{deluxetable}{cccrrrr}		
\tabletypesize{\footnotesize}
\tablewidth{0pt}
\tablecaption{Late B-type Halo Stars in the HVS Survey\label{tab:dat}}
\tablecolumns{7}
\tablehead{
  \colhead{R.A.} & \colhead{Decl.} & \colhead{$g$} & \colhead{$v_{\sun}$} 
  & \colhead{$v_{rf}$} & \colhead{$l$} & \colhead{$b$} \\
  \colhead{(hr)} & \colhead{(deg)} & \colhead{(mag)} & \colhead{(\kms)}
  & \colhead{(\kms)} & \colhead{(deg)} & \colhead{(deg)}
}
	\startdata
 0:01:31.115 & $+$26:46:55.06 & $19.671\pm0.021$ & $-374.4\pm11.8$  & $-178.3$ & 109.35 & $-34.77$ \\
 0:02:33.817 & $-$09:57:06.85 & $18.578\pm0.021$ & $ -87.7\pm~7.7$  & $  -1.8$ &  86.76 & $-69.32$ \\
 0:05:28.141 & $-$11:00:10.07 & $19.271\pm0.042$ & $-115.9\pm~9.2$  & $ -35.5$ &  86.89 & $-70.59$ \\
 0:05:51.194 & $+$32:56:35.75 & $18.075\pm0.020$ & $-250.5\pm~8.7$  & $ -44.9$ & 112.01 & $-28.96$ \\
 0:07:00.445 & $-$19:46:13.52 & $18.304\pm0.029$ & $   8.9\pm~9.7$  & $  55.0$ &  66.44 & $-77.44$ \\
 0:07:57.158 & $-$06:02:46.37 & $18.123\pm0.017$ & $-124.7\pm12.4$  & $ -27.6$ &  94.85 & $-66.52$ \\
 0:12:26.890 & $-$10:47:54.56 & $19.042\pm0.025$ & $-128.2\pm~8.7$  & $ -51.0$ &  91.74 & $-71.27$ \\
 0:13:00.653 & $-$20:23:31.26 & $18.511\pm0.071$ & $ -40.4\pm11.3$  & $   0.1$ &  68.37 & $-78.93$ \\
 0:15:58.669 & $+$33:48:40.41 & $17.494\pm0.018$ & $   8.2\pm11.0$  & $ 210.3$ & 114.56 & $-28.48$ \\
 0:17:22.694 & $+$32:33:54.26 & $19.592\pm0.019$ & $  46.9\pm11.3$  & $ 246.1$ & 114.67 & $-29.76$
	\enddata
\tablecomments{(This table is available in its entirety in machine-readable and
Virtual Observatory forms in the online journal. A portion is shown here for
guidance regarding its form and content.)}
 \end{deluxetable}

\end{document}